\documentclass[12pt,a4paper]{article}
\usepackage{epsfig}
\begin{document}
\title{Lepton flavor violating signals of the neutral top-pion in future lepton colliders}
\author{Chong-Xing Yue, Zheng-Jun Zong, Li Zhou, Shuo Yang\\
{\small Department of Physics, Liaoning  Normal University,
Dalian, 116029. P. R. China}
\thanks{E-mail:cxyue@lnnu.edu.cn}}
\date{\today}
\maketitle

\begin{abstract}
\hspace{5mm} The presence of the top-pions $\pi_{t}^{0,\pm}$ in
the low-energy spectrum  is an inevitable feature of the topcolor
scenario. Taking into account the constraints of the present
experimental limit of the lepton flavor violating($LFV$) process
$\mu\rightarrow e\gamma$ on the free parameters of
topcolor-assisted techicolor(TC2) models, we study the
contributions of the neutral top-pion $\pi^{0}_{t}$ to the $LFV$
processes $\mu^{+}\mu^{-}\rightarrow\tau \mu$ ( or $\tau e$), $
\gamma \gamma\rightarrow\tau \mu$ (or $\tau e$), $e^{+}
e^{-}\rightarrow\tau \mu $, and $e \gamma\rightarrow e
\pi_{t}^{0}\rightarrow e \tau \mu(e)$ via the flavor changing
($FC$) couplings $\pi_{t}^{0}l_{i}l_{j}$ and discuss the
possibility of searching for the $LFV$ signals via these processes
in future lepton colliders.

 \vspace{1cm}


\end {abstract}

\newpage
\noindent{\bf I. Introduction}

The solar neutrino experiments[1] and the atmospheric neutrino
experiments[2] confirmed by reactor and accelerator experiments[3]
have made one believe that neutrinos are massive and oscillate in
flavors, and provide the only direct observation of physics that
cannot be accommodated within the standard model($SM$), which can
be seen as the first experimental clue for the existence of new
physics beyond the $SM$. Thus, the $SM$ requires some modification
to account for the pattern of neutrino mixing, in which the lepton
flavor violating($LFV$) processes are allowed. The observation of
the $LFV$ signals in present or future high-energy experiments
would be a clear signature of new physics beyond the $SM$.

Many kinds of popular specific models beyond the $SM$ predict the
presence of new particles, such as new gauge bosons and new
scalars, which can naturally lead to the tree-level $LFV$
couplings. In general, these new particles could enhance branching
ratios for some $LFV$ processes, and perhaps bringing them into
the observable threshold of the present and next generations of
collider experiments. Furthermore, nonobservability of these $LFV$
processes can lead to strong constraints on the nature of new
physics. Thus, studying the possible $LFV$ signals of new
particles in various high-energy colliders is very interesting and
needed.

To completely avoid the problems arising from the elementary Higgs
field in the $SM$, various kinds of dynamical electroweak symmetry
breaking ($EWSB$) models have been proposed, and among which the
topcolor scenario is attractive because it can explain the large
top quark mass and provide possible $EWSB$ mechanism[4]. Almost
all of these kind of models propose that the scale of the gauge
groups should be flavor nonuniversal. When one writes the
nonuniversal interactions in the masseigen basis, it can induce
the tree -level flavor changing$(FC)$ couplings, which can
generate rich phenomenology.

The presence of the physical top-pions $\pi_{t}^{\ 0,\pm}$ in the
low-energy spectrum is an inevitable feature of the topcolor
scenario, regardless of the dynamics responsible for $EWSB$ and
other quark masses[5]. One of the most interesting features of the
top-pions $\pi_{t}^{\ 0,\pm}$ is that they have large Yukawa
couplings to the third-generation quarks and can induce the
tree-level $FC$ couplings in quark sector and in the lepton
sector. The obviously phenomenological implication of this feature
is that the top-pions can be significant produced via some $FC$
processes, and their possible signatures might be observed in
future hadron colliders[5,6] and lepton colliders[7,8].
Furthermore, the top-pions can generate large corrections to some
observables related with the $FC$ couplings and give
characteristic signatures at various high-energy colliders[9,10].
On the other hand, the top-pions can generate significant
contributions to the $LFV$ processes, such as $l_{i}\rightarrow
l_{j}\gamma$, $l_{i}\rightarrow l_{j}l_{k}l_{l}$ and $Z\rightarrow
l_{i}l_{j}$ [11]. The branching ratios for some of these processes
can be enhanced to their current or future experimental bounds,
which can give severe constraints on the free parameters of
topcolor scenario.

Among the various $LFV$ processes that have been considered in the
literature, the most fruitful ones are the radiative decays
$\tau\rightarrow e\gamma$, $\tau\rightarrow \mu \gamma$,
$\tau\rightarrow e\eta$, and $\mu\rightarrow e\gamma$, since their
branching ratios are tested with high precision[12,13]. These
processes usually provide the restrictive experimental bounds on
the free parameters of the popular specific models beyond the
$SM$. For example, the contributions of the Higgs bosons to these
processes have been extensively studied in Refs.[14,15].  Taking
into account the constraints of the experimental upper limits of
these $LFV$ processes on the free parameters, the $LFV$ decays of
the Higgs bosons and possible signals in the future high-energy
collider experiments have been studied in Ref.[16].

In this paper, we will study the $LFV$ signals of the neutral
top-point $\pi_{t}^{0}$ at the future various lepton colliders in
the context of the topcolor-assisted technicolor $(TC2)$
models[17] with free parameters being compatible with the most
relevant data of $\mu$ radiative decay $\mu\rightarrow e\gamma$.
In this work, we consider the contributions of the neutral
top-pion $\pi_{t}^{0}$ to the $LFV$ processes
$\mu^{+}\mu^{-}\rightarrow\tau \mu$ ( or $\tau e$ ), $\gamma
\gamma\rightarrow\tau \mu$ (or $\tau e$ ), $e^{+}
e^{-}\rightarrow\tau \mu$, and $e \gamma\rightarrow e
\pi_{t}^{0}\rightarrow e \tau \mu(e)$ via the $FC$ couplings
$\pi_{t}^{0}l_{i}l_{j}$ and discuss possibility of directly
searching for the $LFV$ signals in the future lepton colliders via
these processes. For completeness, we also include an estimation
of the contributions of the new gauge boson $Z'$ predicted by
$TC2$ models to these processes and compare them with those of the
neutral top-pion $\pi_{t}^{0}$.

This paper is organized as follows. Section II contains a short
summary of the relevant coupling to ordinary particles and decay
modes of neutral top-pion $\pi_{t}^{0}$ in $TC2$ models.
Discussions of the constraints coming from the $LFV$ $\tau$ and
$\mu$ decays, the muon anomalous magnetic moment $a_{\mu}$, and
other observables on the relevant free parameters are also given
in this section. Section III, IV and V are devoted to the
computation of the produce cross sections generated by
$\pi_{t}^{0}$ exchange for the $LFV$ processes
$\mu^{+}\mu^{-}\rightarrow\tau \mu(\tau e),\gamma
\gamma\rightarrow\tau \mu(\tau e),e^{+} e^{-}\rightarrow\tau
\mu(\tau e)$, and $e \gamma\rightarrow e \tau \mu(e \tau e)$,
respectively. Some phenomenological analyses are also included.
Our conclusions are given in Sec. VI.

\noindent{\bf II. The $LFV$ coupling of $\pi_{t}^{0}$ to ordinary
particles }

For $TC2$ models [17], technicolor$(TC)$ interactions play a main
role in breaking the electroweak symmetry. Topcolor interactions
make small contributions to $EWSB$ and give rise to the main part
of the top quark mass, $(1-\varepsilon)m_{t}$, with the parameter
$\varepsilon<<1$, Thus, there is the following relation:
\begin{equation}
\nu_{\pi}^{2}+F_{t}^{2}=\nu_{w}^{2},
\end{equation}
where $\nu_{\pi}$ represents the contributions of $TC$
interactions to $EWSB$, $\nu_{w}=\nu/\sqrt{2}\simeq$174GeV. Here
$F_{t}\simeq$50GeV is the physical top-pion decay constant. This
means that the masses of the $SM$ gauge bosons $W$ and $Z$ are
given by absorbing the linear combination of the top-pions and
technipions. The orthogonal combination of the top-pion and
technipions remains unabsorbed and physical[18,19,20]. However,
the absorbed Goldstone boson linear combination is mostly the
technipions, while the physical combination is mostly the
top-pions, which are usually called physical
top-pions$(\pi_{t}^{0,\ \pm})$.

For $TC2$ models, the underlying interactions, topcolor
interactions, are nonuniversal and therefore do not possess
Glashow-Iliopoulos-Maiani($GIM$) mechanism. The nonuniversal gauge
interactions result in the new $FC$ coupling vertices when one
writes the interactions in the mass eigenbasis. Thus, the
top-pions can induce the new $FC$ coupling vertices. The couplings
of the neutral top-pion $\pi_{t}^{0}$ to ordinary fermions, which
are related to our calculation, can be written as [5,8,11,17]:
\newpage
\begin{eqnarray}
&&\frac{m_{t}}{\sqrt{2}F_{t}}\frac{\sqrt{\nu_{w}^{2}-F_{t}^{2}}}{\nu_{w}}
     [K_{UR}^{tt}K_{UL}^{tt^{*}}\bar{t}\gamma^{5}t\pi_{t}^{0}+
     \frac{m_{b}-m_{b}'}{m_{t}}\overline{b}\gamma^{5}b\pi_{t}^{0}+K_{UR}^{tc}
     K_{UL}^{tt^{*}}\bar{t_{L}}
     c_{R}\pi_{t}^{0}]\nonumber\\
&&+\frac{m_{l}}{\sqrt{2}\nu_{w}}\bar{l}
     \gamma^{5}l\pi_{t}^{0}+\frac{m_{\tau}}{\sqrt{2}\nu_{w}}
     K_{\tau i}\bar{\tau}\gamma^{5}l_{i}\pi_{t}^{0},
\end{eqnarray}

where $m_{b}'\approx0.1\varepsilon m_{t}$ is the part of the
bottom-quark mass generated by extended technicolor$(ETC)$[9].
$K_{UL}$ and $K_{UR}$ are rotation matrices that diagonalize the
up-quark mass matrix $M_{U}$, i.e. $ K_{UL}^{+} M_{U}
K_{UR}=M_{U}^{dia}$. To yield a realistic form of the
Cabibbo-Kobayashi-Maskawa($CKM$) matrix $V$, it has been shown
that their values can be taken as [5]:
\begin{equation}
K_{UL}^{tt}\approx 1,\ \ \  K_{UR}^{tt}= 1-\varepsilon,\ \ \
K_{UR}^{tc}\leq  \sqrt{2\varepsilon-\varepsilon^{2}}\ .
\end{equation}
In the following calculation, we will take
$K_{UR}^{tc}=\sqrt{2\varepsilon-\varepsilon^{2}}$ and take
$\varepsilon$ as a free parameter, which is assumed to be in the
range of 0.01-0.1[4,17]. $l=\tau,\mu $ or $e$, $l_{i}$(i=1,2) is
the first(second)generation lepton $e$($\mu$), and $k_{\tau i}$ is
the flavor mixing factor between the third-and the first-or
second- generation leptons. Certainly, there is also the $FC$
scalar coupling $\pi_{t}^{0}\mu e$. However, the topcolor
interactions only contact with the third-generation fermions, and
thus, the flavor mixing between the first- and second-generation
fermions is very small, which can be ignored.

The limits on the top-pion mass $m_{\pi_{ t}}$  might be obtained
via studying its effects on various experimental observables. For
example, considering the couplings of the neutral top-pion
$\pi_{t}^{0}$ to bottom quark through instanton effects, Ref.[18]
has shown that the process $b\rightarrow s\gamma$, $B-\bar{B}$
mixing and $D-\bar{D}$ mixing demand that the top-pions are likely
to be light, with masses of the order of a few hundred $GeV$.
Since the negative top-pion corrections to the $Z\rightarrow
b\bar{b}$ branching ratio $R_{b}$ become smaller when the top-pion
is heavier, the electroweak precision measurement data of $R_{b}$
give rise to a certain lower bound on the top-pion mass. It was
shown that the top-pion mass should not be lighter than the order
of $1TeV$ to make TC2 models consistent with the electroweak
precision measurement data[19]. Reference [20] restudied this
problem and found that the top-pion mass $m_{\pi_{t}}$ is allowed
to be in the range of a few hundred $GeV$ depending on the models.
Thus, the value of the top-pion mass $m_{\pi_{t}}$ remains subject
to large uncertainty[4]. In general, the top-pion mass
$m_{\pi_{t}}$ is allowed to be in the range of a few hundred $GeV$
depending on the models. In this paper, we will assume 150$GeV\leq
m_{\pi_{t}}\leq350GeV$. In this case, the possible decay modes of
$\pi_{t}^{0}$ are $b\overline{b},\ \overline{t}c,\
\overline{f}f(f$ is the first-or second-generation fermions, or
the third-generation bottom quark or leptons), $gg,\ \gamma\gamma$
and $\tau l$, ($l=\mu$ or $e$). The total decay width which has
been calculated in Ref.[21]. However, the modes of decay into
leptons are not included in this reference. If we take into
account these decay modes, then the branching ratio
$Br(\pi_{t}^{0}\rightarrow \tau\mu$) is in the range of
$7.2\times10^{-3}k^{2}_{\tau\mu}\sim
2.5\times10^{-5}k^{2}_{\tau\mu}$ for 150$GeV\leq m_{\pi
_{t}}\leq$350$GeV$ and 0.01$\leq\varepsilon\leq 0.1$.

Since the topcolor interactions only contact with the third-
generation fermions and the flavor mixing between the first- and
second-generation fermions is very small, the $ B $ system
observables cannot give significant constraints on the flavor
mixing factor $k_{\tau i}$.

It is well known that the precision measurement of the muon
anomalous magnetic moment $a_{\mu}$ is a sensitive test for the
new physics beyond the $SM$. Comparing the new measurement value
of $a_{\mu}$ with the present $SM$ prediction, there remains a
tantalizing discrepancy[22]:
\begin{equation}
a_{\mu}^{exp}-a_{\mu}^{SM}=(24.5\pm9)\times10^{-10}.
\end{equation}
If we assume that the observed deviation, as shown in $Eq.$(4),
comes from the contributions of new particles predicted by $TC2$
models, then we might obtain a constraint on $TC2$ models using
this deviation[23]. However, we find that the contributions of
$TC2$ models to $a_{\mu}$ come mainly from the $ETC$ gauge bosons
and nonuniversal gauge boson $Z'$; the contributions of
$\pi_{t}^{0}$ are smaller than $1\times 10^{-13}$ i.e. $\delta
a_{\mu}^{\pi_{t}^{0}}<1\times10^{-13}$. Thus, the recently
measurement value of $a_{\mu}$ can not give significant
constraints on the free parameters, which are related to the
top-pion, such as $\varepsilon, m_{\pi _{t}}$, and $k_{\tau_{i}}$.

The neutral top-pion $\pi_{t}^{0}$ can produce significant
contributions to the $LFV$ processes $l_{i}\rightarrow
l_{j}\gamma$, $l_{i}\rightarrow l_{j}l_{k}l_{l}$, and
$Z\rightarrow l_{i}l_{j}$ via the $FC$ couplings
$\pi_{t}^{0}\tau\mu$ and $\pi_{t}^{0}\tau e$, and can enhance the
branching ratios of these processes by several orders of
magnitude[11]. For the processes $l_{i}\rightarrow l_{j}\gamma$,
the contributions come from the on-shell photon penguin diagrams,
while the contributions come from both the on-shell photon penguin
diagrams and the tree-level diagrams for the processes
$\tau\rightarrow l_{i}l_{j}l_{k}$. Reference [11] has shown that,
in all of the parameter space of $TC2$ models, the branching
ratios of the processes $\tau\rightarrow l \gamma$,
$l_{i}\rightarrow l_{j}l_{k}l_{l}$, and $Z\rightarrow l_{i}l_{j}$
are far below the experimental upper bound on these processes,
except for the process $\mu\rightarrow3e$, which might approach
the observable threshold of near-future experiments[13]. The
present experimental limit of the process $\mu\rightarrow e\gamma$
can give severe constraints on the free parameters of $TC2$
models. If we assume $k=k_{\tau\mu}=k_{\tau e}$ and $150GeV\leq
350GeV$, then there must be $k\leq0.16$. Taking into account this
constraint, we will study possible $LFV$ signals of the neutral
top-pion $\pi_{t}^{0}$ at future various lepton colliders in the
following sections.

\noindent{\bf III. LFV signals of the neutral top-pion
$\pi_{t}^{0}$ in the  future muon collders }

A muon collider is an excellent tool to study the properties of a
heavy scalar or pseudoscalar and potential new physics
effects[24]. It has been shown that a large number of Higgs
bosons[24] or new particles, such as technihadron and
technipions[25], can be produced through s-channel resonance
processes. The $FC$ scalar couplings[26,27] might be tested at
future muon colliders. Thus, the future muon collider opens up the
possibility of studying the $LFV$ signals mediated by the new
scalars.

The neutral top-pion $\pi_{t}^{0}$ has $FC$ couplings to the
leptons at tree level and thus can contribute to the processes
$\mu^{+}\mu^{-}\rightarrow \tau\mu(\tau e)$ via $\pi_{t}^{0}$
exchange in the s-channel. In spite of the fact that the coupling
$\pi_{t}^{0}\mu^{+}\mu^{-}$, being proportional to $m_{\mu}/v$, is
very small, if the muon collider with the center-of-mass($c.m.$)
energy $\sqrt{s}$ runs on $\pi_{t}^{0}$ resonance
($\sqrt{s}=m_{\pi_{t}}$), then the neutral top-pion $\pi_{t}^{0}$
may be produced at an appreciable rate[25, 27]. Thus, the neutral
top-pion might give observable $LFV$ signals at future muon
collider experiments.

After averaging over initial polarization and integrating over the
scattering angle, it is straightforward to obtain the unpolarized
cross section of the process $\mu^{+}\mu^{-}\rightarrow
\tau\mu(\tau e)$
 mediated by $\pi_{t}^{0}$ exchange. The s-channel
resonance cross section for this process can be approximately
written as:
\begin{equation}\sigma(\tau\mu(e))\approx\frac{4\pi}{m_{\pi_{t}}}
\cdot Br( \pi_{t}^{0}\rightarrow\mu^{+}\mu^{-})Br(
\pi_{t}^{0}\rightarrow\tau\mu(e)),\end{equation} where
$Br(\pi_{t}^{0}\rightarrow\mu^{+}\mu^{-})$ and $Br(
\pi_{t}^{0}\rightarrow\tau\mu(e))$ represent the branching ratios
of $\pi_{t}^{0}$ decaying to $\mu^{+}\mu^{-}$ and $\tau\mu(e)$,
respectively. Observably, there is $\sigma(\tau \mu)\approx
\sigma(\tau e)$ for $k_{\tau\mu}=k_{\tau e}$ and neglecting the
final state masses, which is different from that generated by the
Higgs bosons[26]. In that case, compared the production cross
section $\sigma(\tau \mu)$, the cross section $\sigma(\tau e)$ is
suppressed by a factor of $m_{e}/m_{\mu}\sim 1/200 $.

\begin{figure}[htb]
\vspace{0 cm}
\begin{center}
\epsfig{file=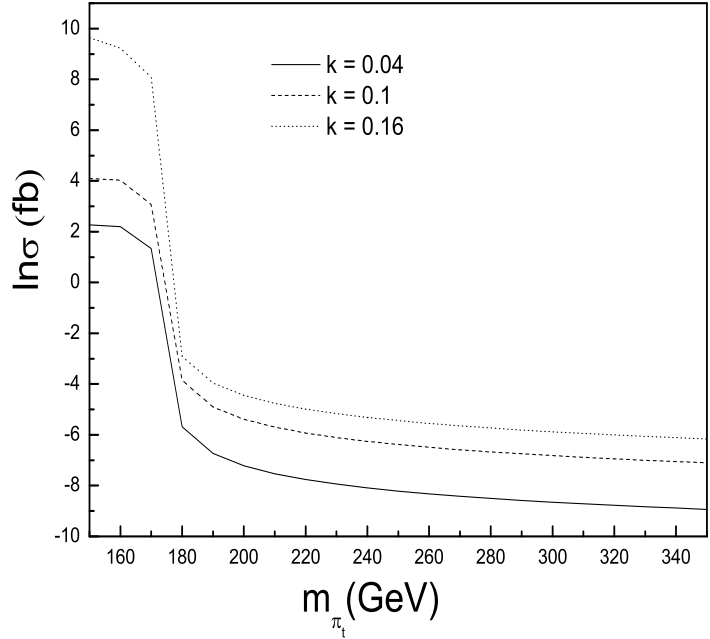,width=300pt,height=280pt} \vspace{-1cm}
\caption{The resonance production cross section $\sigma(\tau\mu)$
as function of the top-pion mass \hspace*{1.8cm} $m_{\pi_{t}}$ for
$\sqrt{s}=m_{\pi_{t}}$ and three values of the mixing parameter
$k$.}
\end{center}
\end{figure}

In $Fig.$1 we show the resonance production cross section
$\sigma(\tau \mu)$ as a function of the top-pion mass
$m_{\pi_{t}}$ for $\sqrt{s}=m_{\pi_{t}}$ and three values of the
mixing parameter $k$. Since the cross section $\sigma(\tau\mu)$ is
not sensitive to the free parameter $\varepsilon$, we have taken
$\varepsilon = 0.05$ in $Fig.$1. One can see from $Fig.$1 that the
$\sigma(\tau \mu)$ decreases with $m_{\pi_{t}}$ increasing and the
mixing parameter $k$ decreasing. For $m_{\pi_{t}}\geq
m_{t}+m_{c}$, the $FC$ channel
$\pi_{t}^{0}\rightarrow\overline{t}c$ opens up and the branching
ratios $Br(\pi_{t}^{0}\rightarrow\mu^{+}\mu^{-})$ and $Br(
\pi_{t}^{0}\rightarrow\tau\mu)$ drop substantially, and thus the
cross sections $\sigma(\tau \mu)$ and $\sigma(\tau e)$ drop
considerably. The values of $\sigma(\tau \mu)$ are in the ranges
of $1.54\times10^{2}\sim3.4\times10^{-2}fb$ and
$5.5\times10^{-2}\sim1.5\times10^{-4}fb$ for $k\leq0.16$,
$150GeV\leq m_{\pi_{t}}<180GeV$, and $180GeV\leq
m_{\pi_{t}}\leq350GeV$, respectively. If we assume a future muon
collider runing with the $c.m.$ energy $\sqrt{s}=200\sim500GeV$
and a yearly integrated luminosity of ${\cal L}=20fb^{-1}$, then
there will be several and up to thousand $\tau\mu$(or $\tau e$)
events to be generated a year for $ m_{\pi_{t}}<180GeV$.

For the $LFV$ processes $\mu^{+}\mu^{-}\rightarrow \tau\mu($ or
$\tau e)$, the final leptons always emerge back to back and
carrying a constant energy which is one half of the $c.m.$ energy
$\sqrt{s}$. The main background would arise from the process
$\mu^{+}\mu^{-}\rightarrow
\overline{\tau}\mu\overline{\nu_{\mu}}\nu_{\tau}$ or
$\overline{\tau}e\overline{\nu_{e}}\nu_{\tau}$. It has been shown
that the contributions to the background come mainly from the
diagrams with Higgs boson exchange[26]. The background cross
section can only reach a peak around $m_{H}=130GeV$ and then drops
quickly out of this range. Thus, as long as $150GeV\leq
m_{\pi_{t}}\leq m_{t}+m_{c}$, the $LFV$ signals of the neutral top
pion $\pi_{t}^{0}$ should be observed via the resonance processes
$\mu^{+}\mu^{-}\rightarrow \tau\mu$(or $\tau e$) in the future
muon colliders.

$TC2$ models also predict the existence of a nonuniversal $U(1)$
gauge boson $Z'$, which can lead to the tree-level $FC$ couplings
$Z'\tau\mu$ and $Z'\tau e$. Thus, the nonuniversal gauge boson
$Z'$ has contributions to the $LFV$ processes
$\mu^{+}\mu^{-}\rightarrow \tau\mu($ or $\tau e)$ via the
s-channel $Z'$ exchange. At leading order, the unpolarized cross
section $\sigma'(\tau\mu)$ generated by $Z'$ exchange can be
written as:
\begin{equation}
\sigma^{'}(\tau\mu)=\frac{25\pi^{2}\alpha^{2}\lambda_{\tau\mu}^{2}}
{12C_{W}^{6}K_{1}}\frac{s}{(s-M_{Z'}^{2})^{2}+M_{Z'}^{2}\Gamma_{Z'}^{2}},
\end{equation}
where $C_{W}=\cos\theta_{W}$, $\theta_{W}$ is the Weinberg angle,
$\lambda_{\tau\mu}$ is the coupling constant of the $FC$ vertex
$Z'\tau\mu$. $K_{1}$ is the mixing parameter between the
nonuniversal gauge boson $Z'$ and the $SM$ gauge boson $Z$. Using
$Eq.(6)$, we can easily give the numerical results. However,
considering the constraints of the electroweak precision
measurement data on the $Z'$ mass $M_{Z'}(M_{Z'}>1TeV)$[4,28], the
cross section $\sigma'(\tau\mu)$ is smaller than $10^{-4}fb$ in
all of the parameter space at the future muon colliders with
$\sqrt{s}=300\sim500GeV$, which cannot produce observable signals.
Furthermore, even if the nonuniversal gauge boson $Z'$ can produce
observable $LFV$ signals, we can use polarized beams to separate
the contributions from $\pi_{t}^{0}$ exchange from those from $Z'$
exchange.

\noindent{\bf IV. The neutral top-pion $\pi_{t}^{0}$ and the $LFV$
processes $\gamma\gamma\rightarrow\pi_{t}^{0}\rightarrow
\tau\mu(\tau e)$ }

It is widely believed that hadron colliders, such as Tevatron and
future $LHC$, can directly probe possible new physics beyond the
$SM$ up to a few $TeV$, while the future high-energy linear
$e^{+}e^{-}$ collider $(ILC)$ is also required to complement the
probe of the new particles with detailed measurement. A future $I
LC$ will offer an excellent opportunity to study production and
decay of the new particles with percent-level-precision[29]. A
unique feature of the future $ILC$ is that it can be transformed
to $\gamma\gamma$ colliders with the photon beams generated by the
Compton backward scattering of the initial electron and laser
beams. Their effective luminosity and energy are expected to be
comparable to those of the $ILC$. In some scenarios, they are the
best instrument for the discovery of signals of new physics[30].

\begin{figure}[htb]
\vspace{0cm}
\begin{center}
\epsfig{file=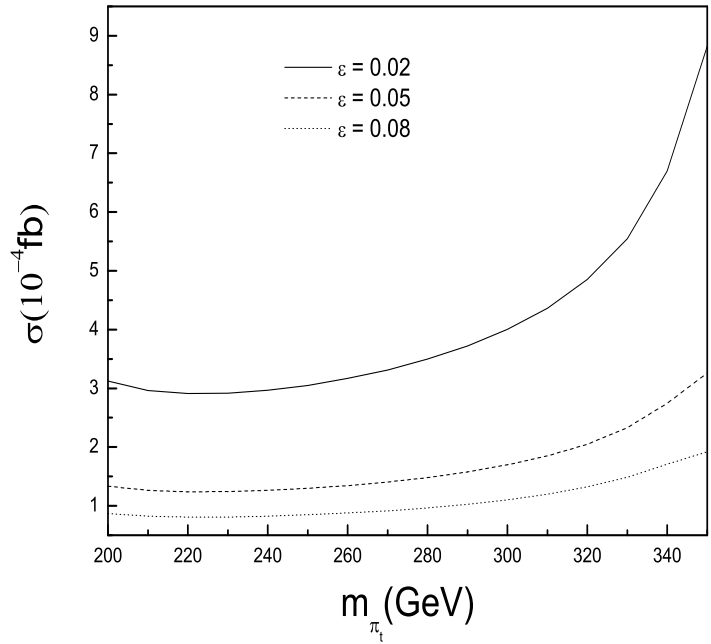,width=300pt,height=280pt} \vspace{-1cm}
\caption{The effective cross section $\sigma(s)$ as a function of
$m_{\pi_{t}}$ for $\sqrt{s}=500GeV, k=0.1$, \hspace*{1.8cm} and
three values of the free parameter $\varepsilon$.}
\end{center}
\end{figure}

Similar to the contributions of the neutral top-pion $\pi_{t}^{0}$
to top-charm associated production at $\gamma\gamma$
colliders[7,9], the contributions of $\pi_{t}^{0}$ to the $LFV$
processes $\gamma\gamma\rightarrow\tau\mu(\tau e)$ via the $FC$
couplings $\pi_{t}^{0}l_{i}l_{j}$ proceed through the self-energy
diagrams, vertex diagrams and the s-channel triangle diagram. We
have explicitly calculated the contributions of these Feynman
diagrams and found that the dominant contributions come from the
s-channel triangle diagram, whereas the remaining diagrams give
negligible contributions. Our numerical results are shown in
$Fig.2$, in which we plot the effective cross section
$\sigma(s)=\sigma(e^{+}
e^{-}\rightarrow\gamma\gamma\rightarrow\tau \mu)$ for the $LFV$
process $\gamma\gamma\rightarrow\tau\mu$ as a function of the
top-pion mass $m_{\pi_{t}}$ for the $c.m.$ energy
$\sqrt{s}=500GeV$, the flavor mixing factor $k=0.1$, and three
values of the parameter $\varepsilon$. From this figure, one can
see that the cross section $\sigma(s)$ for the process
$e^{+}e^{-}\rightarrow\gamma\gamma\rightarrow\tau\mu$ (or $\tau
e$) is relatively insensitive to the parameter $\varepsilon$. As
the top-pion mass $m_{\pi_{t}}$ increases, the cross section
increases monotonically. For $k=0.1$ and $200GeV\leq
m_{\pi_{t}}\leq350GeV$, the value of the cross section
$\sigma(e^{+} e^{-}\rightarrow\gamma\gamma\rightarrow\tau \mu)$ is
smaller than $1\times10^{-3}fb$. Certainly, if we assume
$150GeV\leq m_{\pi_{t}}<180GeV$, then the cross section increases
rapidly and its value can reach $8.3\times10^{-2}fb$.

The flavor mixing parameter $k$ controls the strength of the $FC$
coupling $\pi_{t}^{0}\tau l(l=\mu$ or $e)$ and further determines
the $LFV$ signals of the neutral top-pion $\pi_{t}^{0}$. Its value
is severely constrained by the current experimental upper bound on
the $LFV$ process $\mu\rightarrow e\gamma$[11]. There is $k\leq
0.16$ for $m_{\pi_{t}}\leq 350GeV$. Taking into account this
constraint, we plot the effective cross section $\sigma(s)$ as a
function of $k$ for $\sqrt{s}=500GeV,\ \varepsilon=0.05$, and
three values of the top-pion mass $m_{\pi_{t}}$ in $Fig.$3.

\begin{figure}[htb]
\vspace{-0.5cm}
\begin{center}
\epsfig{file=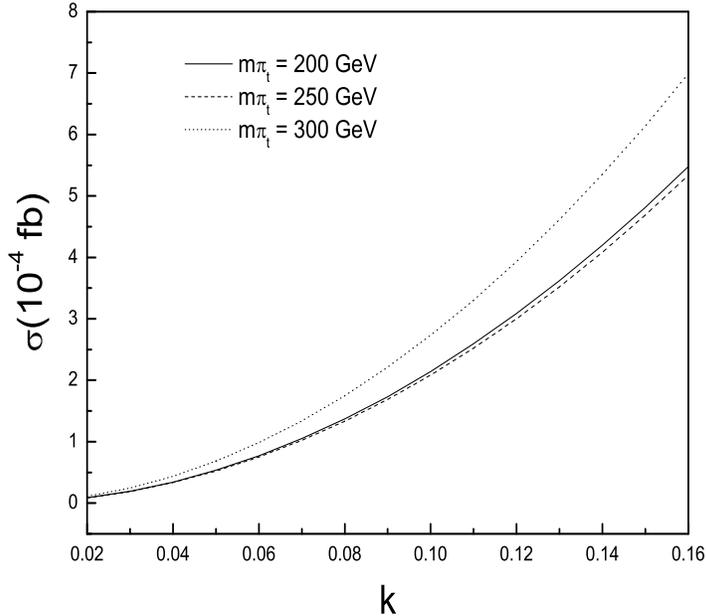,width=300pt,height=280pt} \vspace{-1.5cm}
\caption{The effective cross section $\sigma(s)$ as a function of
$k$ for $\sqrt{s}=500GeV,\ \varepsilon=0.05$ \hspace*{1.8cm}and
three values of the $\pi_{t}^{0}$ mass $m_{\pi_{t}}$.}
\end{center}
\end{figure}

From the above discussion, we can see that, if we assume
$200GeV\leq m_{\pi_{t}}<350GeV$, then the effective cross section
$\sigma(s)$ is smaller than $1\times 10^{-3}fb$ in most of the
parameter space consistent with the constraint from the $LFV$
process $\mu\rightarrow e\gamma$. However, for $150GeV\leq
m_{\pi_{t}}<180GeV$, $\sigma(s)$ is in the range of
$1\times10^{-2}fb\sim8\times10^{-2}fb$. In this case, there will
be several tens $\tau\mu(e)$ events to be generated a year at the
future $ILC$ with $\sqrt{s}=500GeV$ and the yearly integrated
${\cal L}_{int}=340fb^{-1}$, which might be observed in future
$ILC$ experiments.

\begin{figure}[htb]
\vspace{-3.5cm}
\begin{center}
\epsfig{file=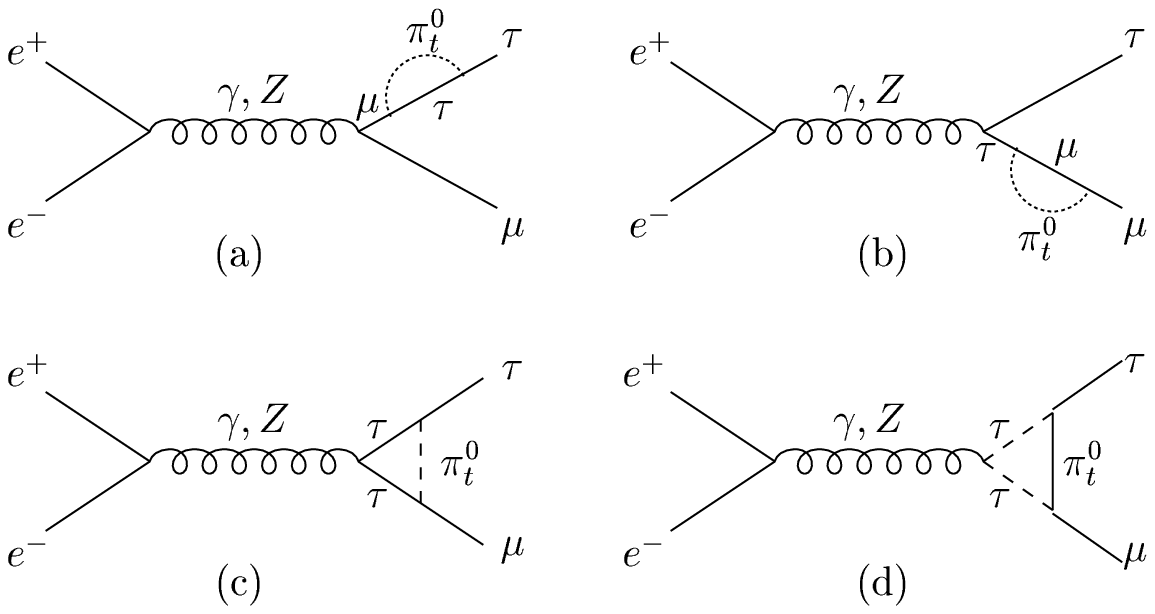,width=600pt,height=900pt} \vspace{-21.5cm}
\caption{Feynman diagrams for the process $e^{+}
e^{-}\rightarrow\tau \mu $ contributed by the $LFV$ coupling
 \hspace*{1.8cm}$\pi_{t}^{0}\tau\mu$}
\end{center}
\end{figure}

The neutral top-pion $\pi_{t}^{0}$ can also contribute to the
$LFV$ processes $e^{+}e^{-}\rightarrow\tau\mu(\tau e)$ via the
$FC$ couplings $\pi_{t}^{0}\tau \mu(e)$. The relevant Feynman
diagrams are shown in $Fig.$4. Since the s-channel process
$e^{+}e^{-}\rightarrow\gamma^{*},\ Z^{*}\rightarrow\tau\mu$ is
suppressed by the propagators of the intermediate photon or
Z-boson in the future $ILC$ with high center-of-mass energy, the
cross section of the process $e^{+}e^{-}\rightarrow\tau\mu$
generated by the neutral top-pion $\pi_{t}^{0}$ should be smaller
than that of the process
$e^{+}e^{-}\rightarrow\gamma\gamma\rightarrow\tau\mu$. We have
confirmed this expectation through explicit calculation. Our
numerical results show that the cross section for the process
$e^{+}e^{-}\rightarrow\tau\mu$ at the $ILC$ with $\sqrt{s}=500GeV$
is smaller than $1\times10^{-7}fb$ in all of the parameter space
of $TC2$ models, which can not be detected in future $ILC$
experiments.

The production cross section of the process $e^{+}e^{-}\rightarrow
Z'\rightarrow \tau\mu(\tau e)$ is approximately equal to that of
the process $\mu^{+}\mu^{-}\rightarrow Z'\rightarrow \tau\mu(\tau
e)$, which is smaller than $1\times10^{-4}fb$ in all of the
parameter space of the $TC2$ models.

\noindent{\bf V. The neutral top-pion $\pi_{t}^{0}$ and the $LFV$
process $e^{-}\gamma\rightarrow e^{-}\tau\mu$ }

A future $ILC$ can also operate in $e^{-}\gamma$ collisions, where
the $\gamma$ beam is generated by the Compton backward scattering
of the incident position and laser beam and its energy and
luminosity can reach the same order of magnitude of the
corresponding position beam[30]. The $e^{-}\gamma$ collisions can
produce particles which are kinematically not accessible at the
$e^{+}e^{-}$ collisions at the same colliders. Thus, the
$e^{-}\gamma$ collisions are well suite for studying the
production and decays of new particles. In this section, we will
discuss the contributions of the neutral top-pion $\pi_{t}^{0}$ to
the $LFV$ process $e^{-}\gamma\rightarrow e^{-}\tau\mu$ and see
whether the $LFV$ signals of $\pi_{t}^{0}$ can be detected via
$e^{-}\gamma$ collisions in the future $ILC$.

\begin{figure}[htb]
\vspace{-9cm}
\begin{center}
\epsfig{file=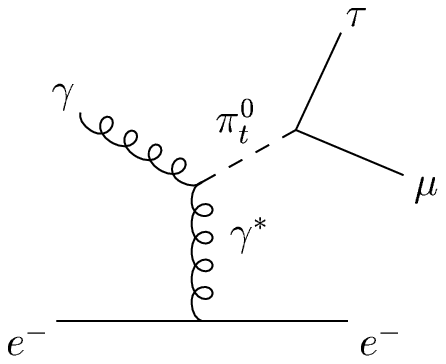,width=780pt,height=940pt} \vspace{-20cm}
\caption{Feynman diagram for the process $e^{-}\gamma\rightarrow
e^{-}\tau\mu$ contributed by the $LFV$ \hspace*{2.0cm}coupling
$\pi_{t}^{0}\tau\mu$}
\end{center}
\end{figure}

In the $TC2$ models, $e^{-}\tau\mu$ production in $e^{-}\gamma$
collisions proceeds through the process $e^{-}\gamma\rightarrow
e^{-}\gamma^{*}\gamma\rightarrow e^{-}\pi_{t}^{0}\rightarrow
e^{-}\tau\mu$, in which the $\gamma$ beam is generated by the
backward Compton scatting of incident positron and laser beam and
the $\gamma^{*}$ beam is radiated from the $e^{-}$ beam. The
relevant Feynman diagram is shown in $Fig.$5. In our calculation,
we use the $Weizs\ddot{a}clcer-Williams$ approximation[31] and
treat the virtual photon $\gamma^{*}$ coming from the $e^{-}$ beam
as a real photon. In this case, the effective cross section of the
$LFV$ process $e^{-}\gamma\rightarrow e\tau\mu$ in the $ILC$
experiments can be obtained by folding the cross section of the
subprocess $\gamma^{*}\gamma\rightarrow \tau\mu$ with the
backscattered photon distribution function $F_{\gamma/e}(x)$[32]
and the function $P_{\gamma/e}(x,E_{e})$, which is the probability
of finding a photon with a fraction $x$ of energy $E_{e}$ in an
ultrarelativistic electron[31].

\begin{figure}[htb]
\vspace{-0.5cm}
\begin{center}
\epsfig{file=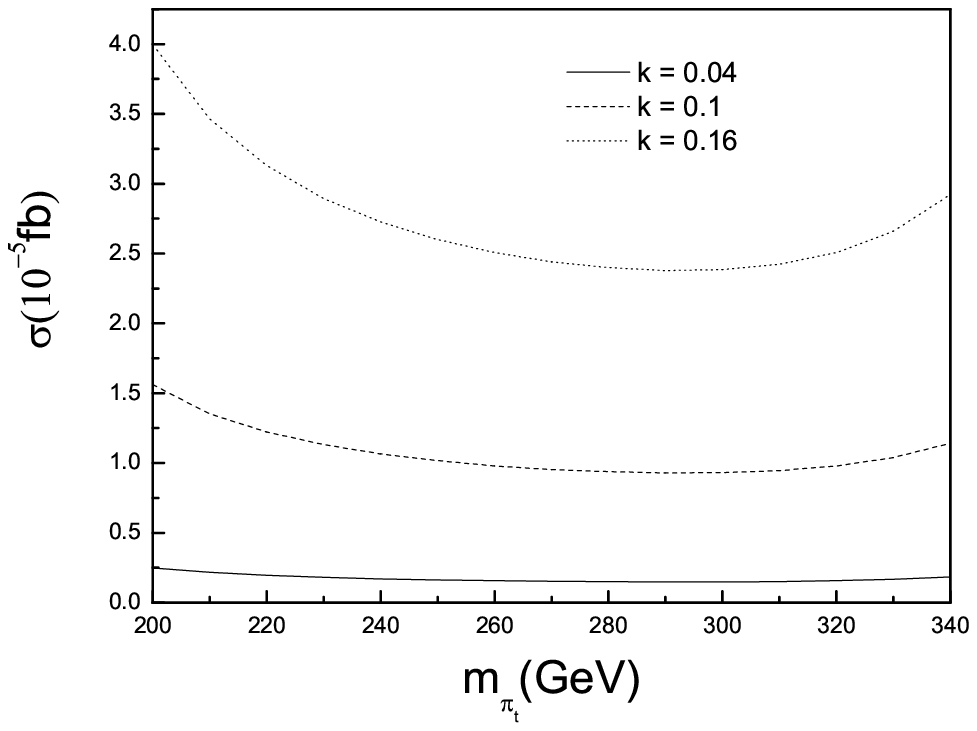,width=300pt,height=280pt} \vspace{-1cm}
\caption{The effective cross section $\sigma(s)$ as a function of
$m_{\pi_{t}}$ for $\sqrt{s}=500GeV$, $\varepsilon=0.05$
\hspace*{1.8cm}and three values of the mixing parameter $k$.}
\end{center}
\end{figure}

Our numerical results are shown in $Fig.$6, in which we plot the
cross section $\sigma(s)$ for the $LFV$ process
$e^{+}e^{-}\rightarrow e^{-}\gamma\rightarrow
e^{-}\gamma\gamma^{*}\rightarrow e^{-}\tau\mu$ as a function of
$m_{\pi_{t}}$ for $\sqrt{s}=500GeV$ and three values of the mixing
parameter $k$. Since the cross section $\sigma(s)$ is not
sensitive to the parameter $\varepsilon$, we have assumed
$\varepsilon=0.05$ in $Fig.$6. One can see from $Fig.$6 that, for
$180GeV\leq m_{\pi_{t}}\leq350GeV$, the value of the cross section
$\sigma(s)$ is smaller than $4\times10^{-5}fb$. Similar to the
case for the process
$e^{+}e^{-}\rightarrow\gamma\gamma\rightarrow\tau\mu$, the
production cross section of this process for $150GeV\leq
m_{\pi_{t}}<180GeV$ is significantly larger than that for
$180GeV\leq m_{\pi_{t}}\leq350GeV$. For $150GeV\leq
m_{\pi_{t}}<180GeV$, the effective cross section of the process
$e^{-}\gamma\rightarrow e^{-}\tau\mu$ can reach
$4.1\times10^{-3}fb$. Even in this case, i.e. $150GeV\leq
m_{\pi_{t}}<180GeV$, the neuatral top-pion $\pi_{t}^{0}$ cannot
produce observable signals via the $LFV$ process
$e^{-}\gamma\rightarrow e^{-}\tau\mu$ in future lepton collider
experiments.

\noindent{\bf VI. Conclusions}

The individual lepton numbers $L_{e}$, $L_{\mu}$, and $L_{\tau}$
are automatically conserved and the tree-level $LFV$ processes are
absent in the $SM$, due to unitary of the leptonic analog of $CKM$
mixing matrix and the masslessness of the three neutrinos.
However, the neutrino oscillation data provide very strong
evidence for mixing and oscillation of the flavor neutrinos, which
imply that the separated lepton numbers are not conserved. Thus,
any observation of the effects for the $LFV$ processes would be a
clear signature of new physics. This fact and the improvement of
the relevant experimental measurements have brought considerable
attention to study these processes in the context of specific
popular models beyond the $SM$ and see whether the $LFV$ effects
can be tested in future high-energy experiments.

The topcolor scenario is one of the important candidates for the
mechanism of $EWSB$. The presence of physical top-pions in the
low-energy spectrum is a common feature of topcolor models. Since
topcolor interactions are assumed to couple preferentially to the
third generation and thus do not possess $GIM$ mechanism, the
physical top-pions have large Yukawa couplings to the third family
fermions and can induce the $FC$ scalar couplings, which might
give significant contributions to the $FC$ processes. The effects
of the top-pion on these processes are governed by its mass
$m_{\pi_{t}}$ and the relevant flavor mixing factors, which might
produce observable signals at future high-energy experiments.

In this paper, we have calculated the contributions of the neutral
top-pion $\pi_{t}^{0}$ predicted by $TC2$ models to the $LFV$
processes $\mu^{+}\mu^{-}\rightarrow\tau \mu(\tau e),\gamma
\gamma\rightarrow\tau \mu(\tau e),e^{+} e^{-}\rightarrow\tau
\mu(\tau e)$, and $e \gamma\rightarrow e \tau \mu(e\tau e)$, and
discussed its possible $LFV$ signals in the future lepton
colliders. We find that the value of the cross sections for these
$LFV$ processes can indeed be enhanced by several orders of
magnitude. With reasonable values of the free parameter, some of
these processes may be within the observable threshold of
near-future lepton collider experiments. For example, taking into
account the constrains of the present experimental limit of the
$LFV$ process $\mu\rightarrow e\gamma$ on the mixing factor $k$
and assuming $150GeV\leq m_{\pi_{t}}<180GeV$, we find that the
cross sections of the $LFV$ processes
$\mu^{+}\mu^{-}\rightarrow\tau \mu$ and
 $e^{+} e^{-}\rightarrow\gamma \gamma\rightarrow\tau \mu$ can
 reach $1.5\times10^{2}fb$ and $8.3\times10^{-2}fb$,
 respectively. However, for $m_{\pi_{t}}\geq200GeV$, the cross
 sections of all of the $LFV$ processes are very small, which cannot be detected
 in future experiments. Thus, we expect that the light
 top-pions predicted by topcolor scenario might produce observable
 $LFV$ signals in future lepton collider experiments.

\vspace{0.5cm} \noindent{\bf ACKNOWLEDGMENTS}

This work was supported in part by Program for New Century
Excellent Talents in University(NCET-04-0209), the National
Natural Science Foundation of China under the Grants No.90203005
and No.10475037, and the Natural Science Foundation of the
Liaoning Scientific Committee(20032101).

\end{document}